# Crossover between Solid-like and Liquid-like Behavior in Supercooled Liquids

X. R. Tian[a*], D. M. Zhang[b*], B. Zhang[a], D. Y. Sun[a§] and X. G. Gong[b§]

[a]School of Physics and Electronic Science, East China Normal University, 200241 Shanghai, China

[b]Key Laboratory for Computational Physical Sciences (MOE), Institute of Computational Physics, Fudan University, Shanghai 200433, China

## Abstract

In supercooled liquids, at a temperature between the glass transition temperature ($T_g$) and the melting point ($T_m$), thermodynamic properties remain continuous, while dynamic behavior exhibits anomalies. The origin of such thermodynamics-dynamic decoupling has long been a puzzle in the field of glass researches. In this study, we show that this feature can effectively characterize by the ratio of the characteristic relaxation time associated with the relative and center-of-mass coordinate of nearest-neighbor atomic pairs. And supercooled liquids can be categorized into two distinct "states" based on their dynamics: solid-like and liquid-like behaviors. We further propose four possible paths from the liquid to the final glass state, each exhibiting unique thermodynamic and dynamic behaviors. Two of these paths predict a characteristic temperature $T_x$ between $T_m$ and $T_g$, where a crossover between solid-like and liquid-like behaviors occurs in supercooled liquids. The molecular dynamics simulations of several supercooled liquids reveal that the actual path followed by all these systems undergo the crossover between solid-like and liquid-like behaviors. $T_x$ is found to reside in a similar temperature range as the critical temperature ($T_c$) in the mode-coupling theory and the breakdown temperature ($T_b$) of the Stokes-Einstein relation. This crossover provides a new microscopic perspective for explaining macroscopic dynamic anomalies, and the absence of a typical thermodynamic phase transition at $T_g$.

[*]These authors contributed equally to this work

[§]Corresponding Authors: xggong@fudan.edu.cn; dysun@phy.ecnu.edu.cn

**Introduction:** At a certain characteristic temperature between the glass transition temperature $T_g$ and the melting point $T_m$, supercooled liquids maintain thermodynamic continuity while exhibiting dynamic anomalous. This kind of anomalous has been suggested in many experiments, e.g., super-exponential increases in dynamic behaviors (or relaxation processes) [1-8], the fragile-to-strong crossover [9-19], and the significant enhancement of dynamic heterogeneity [20-35]. This dynamic-thermodynamic decoupling has remained a central puzzle in understanding the nature of the glass transition [4].

Many theoretical models have predicted this kind of dynamic anomalies around this characteristic temperature. In Goldstein's potential energy landscape (PEL) picture, this characteristic temperature represents the temperature where non-activated dynamics transitions to activated dynamics[36, 37]. Subsequent entropy-energy models[38], the random first order transition theory[39] and spacetime thermodynamic theories[2] further elaborate on this idea, proposing a nonequilibrium first-order phase transition at this characteristic temperature from ergodic to non-ergodic states. The mode-coupling theory (MCT) attributes dynamic anomalies to enhanced interactions between density fluctuation modes, which leads to the dynamic blocking of the system around this characteristic temperature [40].

To capture dynamic anomalies around this characteristic temperature, the previous models usually require an assumption about the nature of the dynamics in supercooled liquids. Such presumed motions are either not well-defined or lack direct evidence. Thus, the complete support from the microscopic perspective still requires further exploration. However, in both theoretical and experimental work, it appears that this characteristic temperature definitely exists. This suggests that near characteristic temperature, supercooled liquids seem to undergo some more fundamental changes in the "state" of supercooled liquids.

In this paper, we identify a crossover temperature ($T_x$) in supercooled liquids, marking a transition from liquid-like to solid-like behavior. This is achieved without presuming specific types or features of atomic motions, relying instead on two fundamental facts and one thermodynamic criterion. The transition may explain the origin of dynamic anomalies in supercooled liquids.

**Theoretical Insight:** Our theoretical analysis is based on two fundamental facts and one thermodynamic criterion. The two fundamental facts are: 1) The system will be in a well-defined liquid state at sufficiently high temperatures and in a solid state at sufficiently low temperatures; 2) The binding strength between atoms always decreases with increasing temperature. The thermodynamic criterion is that in the same phase or state, thermodynamic quantities will continuously change with a definite trend; a change in trends indicates a phase transition or state transition. Based on above facts and criterion, we define a physical quantity $\gamma_\tau$, which shows opposite kinetic trends with temperature in the solid and liquid sides. This quantity $\gamma_\tau$ offers a way to distinguish the dynamic behavior of supercooled liquids, as discussed below:

The $\alpha$-relaxation process of supercooled liquids is usually characterized by the self-intermediate scattering function (SISF, $F(q = q_{max}, t)$) $\langle F(q = q_{max}, t) \rangle = \langle \exp\{i\boldsymbol{q} \cdot [\boldsymbol{r}_i(0) - \boldsymbol{r}_i(t)]\} \rangle$, where $\langle \rangle$ denotes the double average over atoms and ensembles (the same below), and $q_{max}$ is the wave vector associated with the position of the first peak of the static structure factor. Usually, the $\alpha$-relaxation time ($\tau_\alpha$) is obtained from SISF, which corresponds to the time required for the SISF to decay to $e^{-1}$. Considering a nearest-neighbor atomic pair with atomic indices $i$ and $j$, whose center-of-mass coordinate is $\boldsymbol{r}_+(t) = \frac{1}{2}[\boldsymbol{r}_i(t) + \boldsymbol{r}_j(t)]$ and relative coordinate is $\boldsymbol{r}_-(t) = \frac{1}{2}[\boldsymbol{r}_i(t) - \boldsymbol{r}_j(t)]$, the SISF associated with the motion of the center-of-mass (MCM) and the motion of relative coordinates (MRC) of nearest-neighbor atomic pairs labeled as $F_+$ and $F_-$, respectively, can be defined as

$$\langle F_\pm(q = q_{max}, t) \rangle = \langle \sqrt{F_i(q, t) F_j(\pm q, t)} \rangle. \tag{1}$$

Usually, the $\alpha$-relaxation time is obtained from SISF, which corresponds to the time required for the SISF to decay to $e^{-1}$. Using the same mathematical rule in determining $\tau_\alpha$, we can define a characteristic relaxation time $\tau_+$ and $\tau_-$ associated to $F_+$ and $F_-$, respectively. It can be seen that, in a structure relaxation process, $F_+$ ($F_-$) describes the degree of correlation of the parallel (opposite) motion between the nearest-neighbor atomic pair ($i$-th and $j$-th atoms). The ratio $\gamma_\tau = \frac{\tau_+}{\tau_-}$ show opposite trends on the solid and liquid sides (see below). Thus, by analyzing the variation of $\gamma_\tau$ with temperatures, we can qualitatively analyze the state or feature of supercooled liquids.

At much higher temperature, the atomic bonds are very weak, which results in the weak correlation between atomic motions. In this case, $F_i(q,t) \sim F_j(-q,t)$, $\tau_+$ and $\tau_-$ will be very close according to Eq. (1), so $\gamma_\tau$ will tend to one. As the temperature decreases, the bond between neighboring atoms begins to strengthen, thus the relaxation through MRCs becomes more difficult. In contrast, the MCM is less affected. Therefore, with the temperature decreasing, $\gamma_\tau$ gradually decreases. As the temperature continues to decrease, the bond between neighboring atoms strengthens further, $\gamma_\tau$ will continue to decrease until some phase transition or crossover occurs.

For low-temperature glasses, atoms are fixed in certain positions, and $\alpha$-relaxation is almost forbidden. For both MCM and MRC, the $\tau_+$ and $\tau_-$ tend to infinity. In this case, $\gamma_\tau$ also tends to one. As the temperature rising, the bonding strength between atoms weakens, accordingly a small number of atoms will break the constraints of their neighbors and begin to diffuse. As long as one atom in a nearest-neighbor atomic pair is fixed (ignoring thermal vibrations), one of $F_i(\pm q, t)$ and $F_j(\pm q, t)$ will always be equal to 1 (corresponding to the fixed atom). In this case, according to Eq. (1), $F_+(q,t) = F_-(q,t)$, and at this time, $\gamma_\tau \sim 1$. The situation changes when both atoms in a nearest-neighbor atomic pair can diffuse, $\gamma_\tau$ begin to decrease. As the temperature rises further, the system begins to have a certain number of neighboring pairs that can diffuse, $\gamma_\tau$ begins to decrease significantly.

According to above discussions, the variation of $\gamma_\tau$ with temperature is different in mechanism at the liquid side (high temperature) and at the solid side (low temperature). At the liquid side, $\gamma_\tau$ decreases with decreasing temperature, originating from the strengthening of interactions between atoms; whereas at the solid side, $\gamma_\tau$ decreases with increasing temperature, originating from the weakening of interactions between atoms. Based on the variation of $\gamma_\tau$ with temperature, we can divide supercooled liquids into two dynamically different categories. When $\gamma_\tau$ decreases with decreasing temperature, the behavior of supercooled liquids is an extension of liquids, referred to as liquid-like supercooled liquid (LLL); when $\gamma_\tau$ decreases with increasing temperature, the behavior of supercooled liquids likes an extension of solids, referred to as solid-like supercooled liquid (SLL). Note that we do not and need not actually heat the solid glass here. "An extension of solids" refers to the increase of $\gamma_\tau$ of supercooled liquids with decreasing temperature.

According to the trend of $\gamma_\tau$ with temperature, we can infer that there are at least four possible paths from the liquid to the glass transition, as shown in Fig. 1(a)-(d). In Fig. 1, the red and green solid lines correspond to LLL and SLL, respectively, and the horizontal black dotted line corresponds to $\gamma_\tau = 1$. The differences between these four paths are reflected in three aspects: 1) whether the dynamic and thermodynamic behaviors are synchronous; 2) between $T_g$ and $T_m$, the supercooled liquids belong to LLL or/and SLL; 3) how LLL transition to SLL, crossover or jump.

The first path ($P_I$) is shown in Fig. 1(a). In $P_I$, $\gamma_\tau$ undergoes a sudden change at $T_m$, jumping from the red solid line to the green solid line, directly becoming SLL. There is no LLL in supercooled liquids. In this path, the dynamic and thermodynamic behaviors are clearly asynchronous. The dynamic anomalous occurs at $T_m$, while the thermodynamic glass transition occurs at $T_g$. The second possible path $P_{II}$ shown in Fig. 1(b), $\gamma_\tau$ undergoes a sudden change at $T_g$, and supercooled liquids always have the feature of LLL, but without SLL. In this path, the thermodynamic and dynamics anomalous are synchronous.

The third path $P_{III}$ shown in Fig. 1(c), features a characteristic temperature $T_x$ between $T_g$ and $T_m$. Around $T_x$, $\gamma_\tau$ undergoes a sudden jump. In the temperature range above $T_x$ and below $T_m$, the supercooled liquid belongs to LLL. In the temperature range below $T_x$ and above $T_g$, the system exhibits the features of SLL. A transition from LLL to SLL occurs at $T_x$. In this path, the dynamic anomalous occurs at $T_x$, which is clearly asynchronous with the thermodynamic glass transition occurring at $T_g$. If the transition between LLL and SLL is not a jump but a crossover, it becomes the fourth path ($P_{IV}$) shown in Fig. 1(d). The core difference between $P_{III}$ and $P_{IV}$ lies in the transition behavior between LLL and SLL; the former is a jump, while the latter is a crossover.

It should be noted that these four paths are only the most likely ones considered from a physical standpoint. Mathematically, the occurrence of other paths cannot be ruled out. Additionally, the curves in Fig. 1 may change with the cooling rate. For cooling rates above the critical cooling rate, these curves may not undergo qualitative changes. However, for cooling rates below the critical cooling rate, supercooled liquids will crystallize at the freezing temperature. At this point, $\gamma_\tau$ will undergo a sudden change at the crystallization temperature (the horizontal black dotted line in the figure, corresponding to $\gamma_\tau = 1$). Since $\alpha$-relaxation is almost forbidden in crystals, naturally $\gamma_\tau$ equals 1. Next, we will calculate $\gamma_\tau$ for several typical glass-forming liquids to determine the real paths in supercooled liquids.

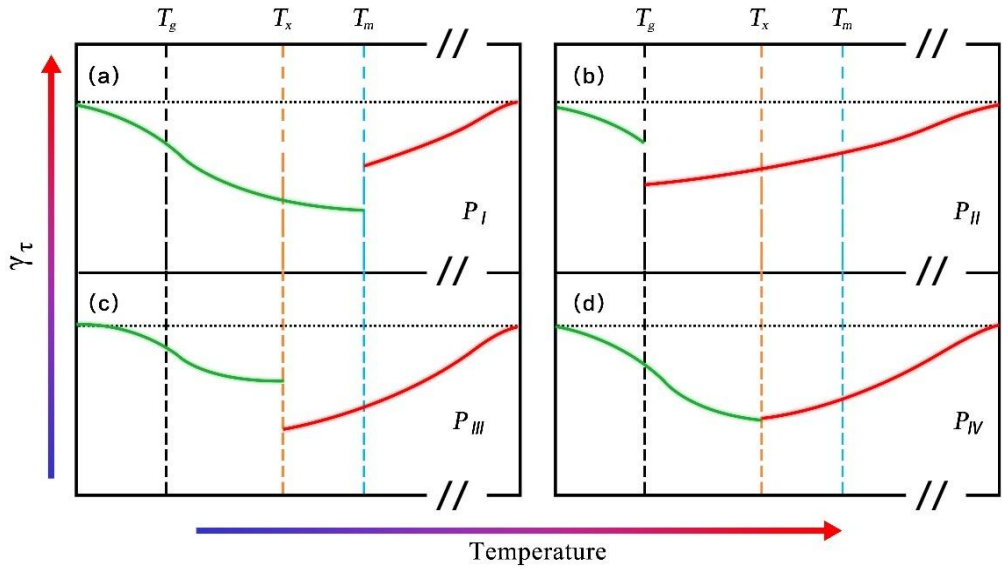

Figure 1. Schematic diagram of the possible paths according to the change in $\gamma_\tau$ with temperature. $T_g$ and $T_m$ are the glass transition temperature and the melting point, respectively. And $T_x$ is a new characteristic temperature predicted in the present work. The horizontal black dotted line in the figure corresponds to $\gamma_\tau = 1$. (a), (b), (c), (d) represent four possible paths ($P_I, P_{II}, P_{III}, P_{IV}$), respectively. The main differences

*between these four paths are threefold: 1) whether the dynamic and thermodynamic behaviors are synchronous; 2) how LLL transition to SLL, crossover or jump.*

**Numerical Results and Discussions:** Aluminum, $Cu_{50}Zr_{50}$, and the well-recognized Kob-Andersen ($N_A:N_B$=80:20) binary Lennard-Jones mixture (BLJ) are investigated in this paper. The interactions between atoms in aluminum, $Cu_{50}Zr_{50}$, and BLJ are taken from reference [41-43], respectively. For the BLJ model, the potential parameters correspond to the $Ni_{80}P_{20}$ alloy [44].

Molecular dynamics (MD) simulations were performed using the LAMMPS code [45]. The simulation cell was a cubic box containing 2048 atoms for Al, $Cu_{50}Zr_{50}$, and 1000 atom for BLJ. The periodic boundary conditions were applied in all three directions. The system was first equilibrated at a well-defined liquid state, and then cooled to low temperatures using the NPT ensemble at a given cooling rate. For each system, more than 30 independent cooling processes were carried out. Each cooling process was completed from a different initial state at the given cooling rate. The cooling rates for $Cu_{50}Zr_{50}$, Al, and BLJ were $1 \times 10^{12}\,K/s$, $1 \times 10^{10}\,K/s$, $9 \times 10^{8}\,K/s$, respectively, all of which were greater than the corresponding critical cooling rates. After the cooling process, at each temperature of interest, long-time canonical ensemble (NVT) simulations were performed, in which the state obtained from NPT simulations was taken as the initial state. Each NVT simulation lasted for 50 ns, and all calculations were performed within this 50 ns. To calculate $\tau_{\pm}$ and $\tau_{\alpha}$, the 50ns MD data is further divided into 5 segments, each approximately 10ns in length. The final $\tau_{\pm}$ and $\tau_{\alpha}$ are the average over all 30×5 MD segments, which is equivalent to the average over more than 30×5 ensembles.

Nearest-neighbor atomic pairs are defined as the atoms whose distance is less than a critical value at the initial moment $t = 0$ of each MD segment. The critical distance corresponds to the position of the first minimum of the pair distribution function, which is 3.71 Å，3.92 Å，1.39 Å for $Cu_{50}Zr_{50}$, Al, and BLJ, respectively. Although these two atoms may no longer be neighbors in subsequent motion, the difference between MRC and MCM can still be well presented in statistical results. Particularly the $\alpha$-relaxation is quite slow in supercooled liquids, such neighboring relationship can be maintained for a considerable length of time. It should be noted that, the length of each MD segment cannot be too long either. Especially, it is essential to avoid that the separated nearest-neighbor atomic pairs re-combinate together in a MD segment. The preceding theoretical analysis has concentrated on how nearest-neighbor atomic pairs tend to separate during the $\alpha$-relaxation process, which is the opposite of the combination process.

MCT predicts the relationship [40] between $T_c$ and $\tau_\alpha$:

$$\tau_\alpha \propto (T - T_c)^{-\nu} \tag{2}$$

However, obtaining $T_c$ through this relationship is not easy, since Eq. (2) is only valid when the temperature is very close to $T_c$. To determine $T_c$, we performed fitting in the range of $\frac{T-T_c}{T_c} < 0.15$. Specifically, we first selected a temperature $T_i$ and then fitted the data in the range $[T_i : \sim 1.15 T_i]$ using Eq. (2) to obtain a $T_c$. Then $T_i$ is moved from low to high temperatures, obtaining a $T_c$ for each testing $T_i$. Finally, $T_c$ is determined with the small fitting error and $T_c$ closest to $T_i$.

In typical liquids, the Stokes-Einstein relation is of great importance as it connects the diffusion coefficient, viscosity, and temperature. However, it was found that this relation breaks down at a specific temperature (denoted as the Stokes-Einstein breakdown temperature $T_b$) [46-50]. Interestingly, $T_b$ is found to be very close to $T_c$ [51-54]. To determine $T_b$, we calculated the self-diffusion coefficient ($D$) of atoms. The self-diffusion coefficient was also determined from the linear relationship by the mean square displacement $\left(\langle R^2(t) \rangle = \langle \left(\boldsymbol{r}_i(0) - \boldsymbol{r}_i(t)\right)^2 \rangle\right)$.

Fig. 2 plots $\tau_+$ and $\tau_-$ as a function of temperature for several supercooled liquids. As expected, $\tau_+$ (red circles) of nearest-neighbor atomic pairs is always shorter than that of $\tau_-$ (blue circles). We found that $\tau_\alpha$ (black circles) is always shorter than $\tau_+$ and $\tau_-$. According to Eq. (1), we have $\langle F_\pm(q = q_{max}, t) \rangle \leq \frac{1}{2}\left(\langle F_i(q,t) \rangle + \langle F_j(\pm q, t) \rangle\right)$. Given that the decay of SISF follows an exponential form, we can obtain

$$\exp\left(-\frac{t}{\tau_\pm}\right) \leq \frac{1}{2}\left(\exp\left(\frac{t}{\tau_i}\right) + \exp\left(\frac{t}{\tau_j}\right)\right) = \exp\left(\frac{t}{\tau_\alpha}\right). \quad (3)$$

Eq. (3) indicates that $\tau_\pm$ is always larger than $\tau_\alpha$. The equality in Eq. (3) holds only when atoms are uncorrelated, which is theoretically only possible in an ideal gas. Any forms of correlations will make $\tau_\pm$ larger than $\tau_\alpha$. Even if the structural information of SISF follows the stretched exponential relaxation as shown in most cases, the above analysis still holds.

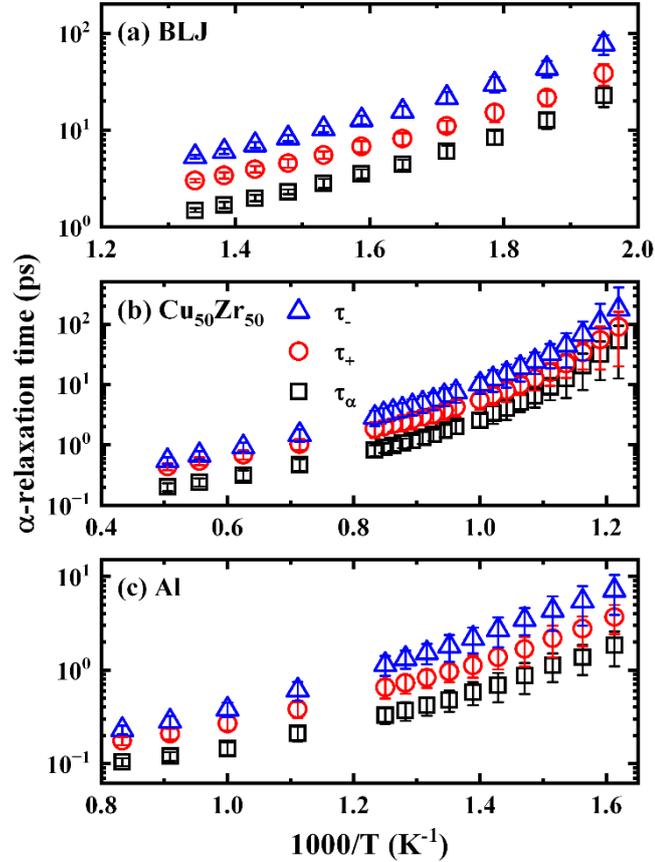

Figure 2. $\tau_+$ and $\tau_-$ as a function of temperature for BLJ (a), $Cu_{50}Zr_{50}$ (b), and Al (c). The red and blue circles represent $\tau_+$ and $\tau_-$, respectively. For comparison, $\tau_\alpha$ is also plotted (black circles).

Based on the variation of $\gamma_\tau$ with temperature, we found that all supercooled liquids studied in this work follow the path $P_{IV}$ during the glass transition, as shown in Fig. 3. Here, dynamics and thermodynamics are asynchronous. From Fig. 3, it can be seen that for all systems, $\gamma_\tau$ gradually decreases from the liquid side, reaching a minimum at a temperature $T_x$ between $T_g$ and $T_m$. As the temperature continues to decrease, $\gamma_\tau$ gradually increases again. Based on the theoretical analysis above, in the temperature range above $T_x$ and below $T_m$, the system exhibits liquid-like behavior, characterized by LLL. In the temperature range from $T_g$ to $T_x$, the supercooled liquid can be considered as a continuation of solid behavior, characterized by SLL.

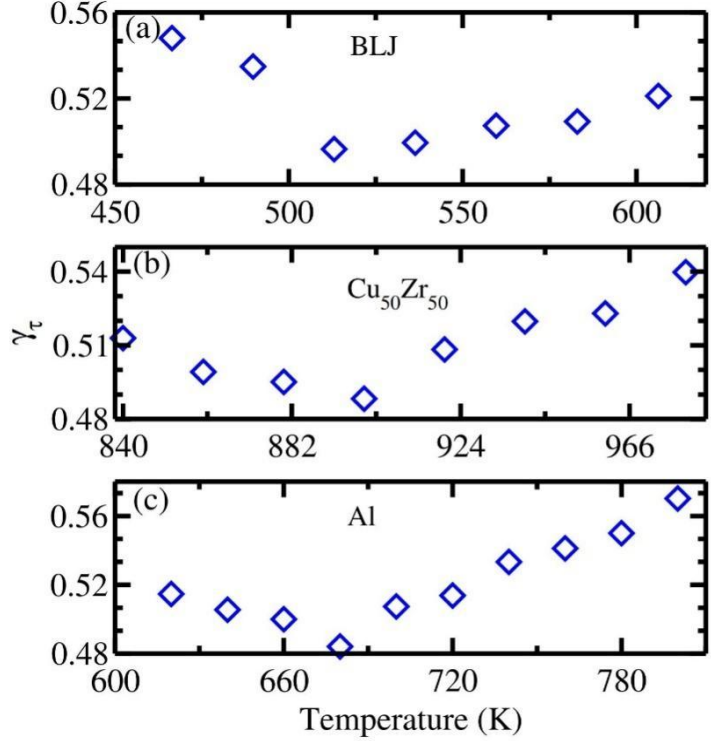

Figure 3. $\gamma_\tau$ as a function of temperature for BLJ (a), $Cu_{50}Zr_{50}$ (b), and Al (c).

The LLL-SLL crossover will definitely cause dynamic anomalies, as the two supercooled liquids are different from the kinetic perspective. The fact, $T_x$ lying between $T_g$ and $T_m$, makes us believe that $T_x$ corresponds the characteristic temperature mentioned in the introduction part. To confirm this issue, we compared $T_x$ with the Stokes-Einstein breakdown temperature $T_b$ and the critical temperature $T_c$ in MCT, which are two typical temperatures related to the dynamic anomalies.

To calculate $T_b$, we plotted the product of the diffusion coefficient and relaxation time as a function of temperature in Fig. 4. If the Stokes-Einstein relationship holds, the product of the diffusion coefficient and relaxation time ($D\tau_\alpha$) should be linear with temperature. From Fig. 4, it can be seen that for all studied systems, when the temperature is above a certain temperature, $D\tau_\alpha$ shows a clear linear relationship with temperature. The dashed lines in the figure are the linear fittings in high temperature region, and it can be seen that the data almost perfectly fit the linear relationship. However, when the temperature is below a certain temperature, the data no longer satisfy the linear relationship, indicating the breakdown of the Stokes-Einstein relation. This temperature is the so-called $T_b$, and the corresponding data are listed in Table 1. From Table 1, it can be seen that for pure substances, $T_x$ and $T_b$ are very close.

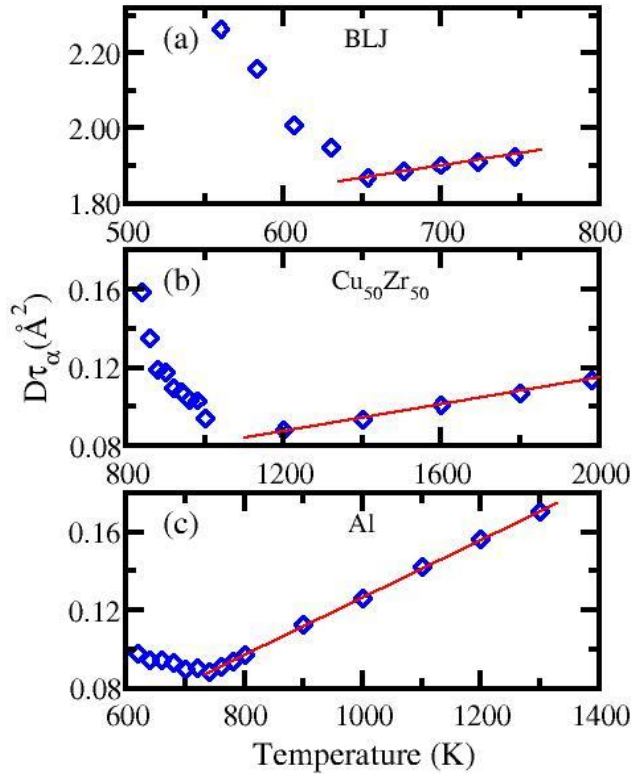

Figure 4. Variation of the product of the diffusion coefficient and relaxation time $D\tau_\alpha$ with temperature for BLJ (a), CuZr (b), and Al (c). If the product is a linear function of temperatures, the Stokes-Einstein relation holds. The dashed lines are linear fits of the high-temperature data.

Previous studies have shown that $T_c$ predicted by MCT is also between $T_g$ and $T_m$. The fact that $T_c$ and $T_b$ are very close is already known[1, 51-53]. This leads us to speculate that $T_c$ and $T_x$ may be the same one. Our results support this speculation, $T_c$ and $T_x$ are indeed very close, as can be seen from Table 1. This speculation is also based on the similarity in the physical nature of the two characteristic temperatures.

Both $T_c$ and $T_x$ reflect some dynamic anomalies. Essentially, both describe a fundamental change in the fluidity of supercooled liquids. The current results show that the physical root of the dynamic anomalies may all originate from a change in the nature of supercooled liquids, that is, from LLL to SLL.

*Table I. Comparison of characteristic temperatures. $T_b$: Stokes-Einstein relations breakdown temperature; $T_c$: MCT critical temperature; $T_x$: the crossover temperature between liquid-like and solid-like behavior. $T_c$ was obtained by fitting in the temperature range greater than $1.2T_g$.*

| System | $T_b$ (K) | $T_c$ (K) | $T_x$ (K) |
|---|---|---|---|
| Al | 720 ± 20 | 645 ± 14 | 680 ± 20 |
| BLJ | 650 ± 23 | 491 ± 8 | 502 ± 23 |
| $Cu_{50}Zr_{50}$ | 1200 ± 20 | 792 ± 9 | 900 ± 20 |

Finally, a few comments on some topics can be addressed: 1) A large number of studies have attempted to identify soft spots or liquid-like atom in glasses or in supercooled liquids [55-61]. SLL provides a logical basis for the existence of soft spots, since only in a solid-like state, so-called liquid-like atoms can be discussed or defined. 2) Compared with the dynamic anomalous, the glass transition around $T_g$ does not exhibit significant thermodynamic phase transition behavior. This is because, near $T_g$, only SLL is involved, therefore discontinuities or divergences in thermodynamic quantities are unlikely. This statement also aligns with the common view that the glassy state can be considered as a frozen supercooled liquid. 3) The supercooled liquids studied in this paper all follow the fourth path ($P_{IV}$). Is it possible that some supercooled liquids follow different paths during glass transitions? We believe this is a very worthwhile question to explore.

**Conclusion:** By theoretically analyzing a characteristic relaxation time associated with the motion of center-of-mass and relative coordinates of nearest-neighbor atomic pairs, we found that there exists two dynamical different supercooled liquid: liquid-like and solid-like one. And a characteristic crossover temperature $T_x$ was identified. Above $T_x$, supercooled liquids exhibit distinct liquid-like characteristics, while below $T_x$, they exhibit solid-like feature. The dynamics anomalies correspond to the competition between the two types of supercooled liquids. This finding demonstrates that during the glass transition from supercooled liquids to glasses, dynamics and thermodynamics are asynchronous. Our finding implies that the Stokes-Einstein relation breakdown and the dynamical arrest in MCT, may have a unified microscopic origin arisen by the dynamical crossover.

**Acknowledgements:** This work was supported by the National Key Research and Development Program of China (Grant No. 2022YFA1404603), and by the National Natural Science Foundation of China (Grant No. 12274127 and 12188101).

**References:**

[1] A. Drozd-Rzoska, S.J. Rzoska, S. Starzonek, New scaling paradigm for dynamics in glass-forming systems, Progress in Materials Science, 134 (2023) 101074.
[2] D. Chandler, J.P. Garrahan, Dynamics on the Way to Forming Glass: Bubbles in Space-Time, Annual Review of Physical Chemistry, 61 (2010) 191-217.
[3] L. Berthier, G. Biroli, Theoretical perspective on the glass transition and amorphous materials, Reviews of Modern Physics, 83 (2011) 587-645.
[4] J.C. Dyre, Colloquium: The glass transition and elastic models of glass-forming liquids, Reviews of Modern Physics, 78 (2006) 953-972.
[5] V. Lubchenko, P.G. Wolynes, Theory of Structural Glasses and Supercooled Liquids, Annual Review of Physical Chemistry, 58 (2007) 235-266.
[6] C.P. Royall, S.R. Williams, The role of local structure in dynamical arrest, Physics Reports, 560 (2015) 1-75.
[7] X.-M. Yang, Q. Yang, T. Zhang, H.-B. Yu, Probing slow glass dynamics down to 10^{-5} Hz, Applied Physics Reviews, 11 (2024) 041403.
[8] Y. Nishikawa, L. Berthier, Collective relaxation dynamics in a three-dimensional lattice glass model, Physical Review Letters, 132 (2024) 067101.
[9] U. Tracht, M. Wilhelm, A. Heuer, H. Feng, K. Schmidt-Rohr, H.W. Spiess, Length Scale of Dynamic Heterogeneities at the Glass Transition Determined by Multidimensional Nuclear Magnetic Resonance, Physical Review Letters, 81 (1998) 2727-2730.
[10] H. Sillescu, Heterogeneity at the glass transition a review, Journal of Non-Crystalline Solids, 243 (1999) 81-108.

[11] M.T. Cicerone, M.D. Ediger, Relaxation of spatially heterogeneous dynamic domains in supercooled ortho-terphenyl, The Journal of Chemical Physics, 103 (1995) 5684-5692.
[12] W. Kob, C. Donati, S.J. Plimpton, P.H. Poole, S.C. Glotzer, Dynamical Heterogeneities in a Supercooled Lennard-Jones Liquid, Physical Review Letters, 79 (1997) 2827-2830.
[13] A.J. Dunleavy, K. Wiesner, R. Yamamoto, C.P. Royall, Mutual information reveals multiple structural relaxation mechanisms in a model glass former, Nature Communications, 6 (2015) 6089.
[14] R.P. White, S. Napolitano, J.E. Lipson, Mechanistic picture for the slow Arrhenius process in glass forming systems: The collective small displacements model, Physical review letters, 134 (2025) 098203.
[15] Y. Hu, F. Li, M. Li, H. Bai, W. Wang, Structural signatures evidenced in dynamic crossover phenomena in metallic glass-forming liquids, Journal of Applied Physics, 119 (2016) 205108.
[16] W. Chu, J. Yu, N. Ren, Z. Wang, L. Hu, A fractal structural feature related to dynamic crossover in metallic glass-forming liquids, Physical Chemistry Chemical Physics, 25 (2023) 4151-4160.
[17] F. Mallamace, C. Branca, C. Corsaro, N. Leone, J. Spooren, S.-H. Chen, H.E. Stanley, Transport properties of glass-forming liquids suggest that dynamic crossover temperature is as important as the glass transition temperature, Proceedings of the National Academy of Sciences, 107 (2010) 22457-22462.
[18] S. Chong, S. Chen, F. Mallamace, A possible scenario for the fragile-to-strong dynamic crossover predicted by the extendedmode-coupling theory for glass transition, Journal of Physics: Condensed Matter, 21 (2009) 504101.
[19] H.C. Andersen, Molecular dynamics studies of heterogeneous dynamics and dynamic crossover in supercooled atomic liquids, Proceedings of the National Academy of Sciences, 102 (2005) 6686-6691.
[20] Kob, Andersen, Testing mode-coupling theory for a supercooled binary Lennard-Jones mixture I: The van Hove correlation function, Physical Review E, 51 (1995) 4626-4641.
[21] E. Flenner, M. Zhang, G. Szamel, Analysis of a growing dynamic length scale in a glass-forming binary hard-sphere mixture, Physical Review E, 83 (2011) 051501.
[22] G. Brambilla, D. El Masri, M. Pierno, L. Berthier, L. Cipelletti, G. Petekidis, A.B. Schofield, Probing the Equilibrium Dynamics of Colloidal Hard Spheres above the Mode-Coupling Glass Transition, Physical Review Letters, 102 (2009) 085703.
[23] G. Biroli, J.P. Bouchaud, A. Cavagna, T.S. Grigera, P. Verrocchio, Thermodynamic signature of growing amorphous order in glass-forming liquids, Nature Physics, 4 (2008) 771-775.
[24] E. Flenner, G. Szamel, Dynamic heterogeneities above and below the mode-coupling temperature: Evidence of a dynamic crossover, J Chem Phys, 138 (2013) 12A523.
[25] K. Kim, S. Saito, Multiple length and time scales of dynamic heterogeneities in model glass-forming liquids: A systematic analysis of multi-point and multi-time correlations, J Chem Phys, 138 (2013) 12A506.
[26] Q. Wang, L.-F. Zhang, Z.-Y. Zhou, H.-B. Yu, Predicting the pathways of string-like motions in metallic glasses via path-featurizing graph neural networks, Science Advances, 10 (2024) eadk2799.
[27] C. Herrero, L. Berthier, Direct numerical analysis of dynamic facilitation in glass-forming liquids, Physical Review Letters, 132 (2024) 258201.
[28] C. Cockrell, R. Grimes, Crossover in atomic mobility underlying the glass transition in inorganic glasses, Journal of Physics: Condensed Matter, 37 (2024) 095402.
[29] S. Berkowicz, I. Andronis, A. Girelli, M. Filianina, M. Bin, K. Nam, M. Shin, M. Kowalewski, T.

Katayama, N. Giovambattista, Supercritical density fluctuations and structural heterogeneity in supercooled water-glycerol microdroplets, Nature Communications, 15 (2024) 1-12.

[30] A. Tahaei, G. Biroli, M. Ozawa, M. Popović, M. Wyart, Scaling description of dynamical heterogeneity and avalanches of relaxation in glass-forming liquids, Physical Review X, 13 (2023) 031034.

[31] Y. Lü, H. Qin, C. Guo, Vortex structure of excitation fields in a supercooled glass-forming liquid and its relationship with relaxations, Physical Review B, 104 (2021) 224103.

[32] L. Berthier, Self-induced heterogeneity in deeply supercooled liquids, Physical Review Letters, 127 (2021) 088002.

[33] M. Ozawa, G. Biroli, Elasticity, facilitation, and dynamic heterogeneity in glass-forming liquids, Physical Review Letters, 130 (2023) 138201.

[34] H. Zhang, Q. Zhang, F. Liu, Y. Han, Anisotropic-isotropic transition of cages at the glass transition, Physical Review Letters, 132 (2024) 078201.

[35] A. Heuer, Exploring the potential energy landscape of glass-forming systems: from inherent structuresvia metabasins to macroscopic transport, Journal of Physics: Condensed Matter, 20 (2008) 373101.

[36] M. Goldstein, Viscous liquids and the glass transition: A potential energy barrier picture, Journal of Chemical Physics, 51 (1969) 3728-3739.

[37] K. Shiraishi, H. Mizuno, A. Ikeda, Johari–Goldstein β relaxation in glassy dynamics originates from two-scale energy landscape, Proceedings of the National Academy of Sciences, 120 (2023) e2215153120.

[38] G. Adam, J.H. Gibbs, On the temperature dependence of cooperative relaxation properties in glass-forming liquids, J Chem Phys, 43 (1965) 139-146.

[39] T.R. Kirkpatrick, D. Thirumalai, P.G. Wolynes, Scaling concepts for the dynamics of viscous liquids near an ideal glassy state, Physical Review A, 40 (1989) 1045-1054.

[40] S.P. Das, Mode-coupling theory and the glass transition in supercooled liquids, REVIEWS OF MODERN PHYSICS, 76 (2004) 785.

[41] F. Ercolessi, J.B. Adams, Interatomic potentials from first-principles calculations: the force-matching method, Europhys Lett, 26 (1994) 583-588.

[42] M.I. Mendelev, Y. Sun, F. Zhang, C.Z. Wang, K.M. Ho, Development of a semi-empirical potential suitable for molecular dynamics simulation of vitrification in Cu-Zr alloys, J Chem Phys, 151 (2019) 214502.

[43] Kob, Andersen, Testing mode-coupling theory for a supercooled binary Lennard-Jones mixture. II. Intermediate scattering function and dynamic susceptibility, Physical review. E, Statistical physics, plasmas, fluids, and related interdisciplinary topics, 52 (1995) 4134-4153.

[44] Weber, Stillinger, Local order and structural transitions in amorphous metal-metalloid alloys, Physical review. B, Condensed matter, 31 (1985) 1954-1963.

[45] A.P. Thompson, H.M. Aktulga, R. Berger, D.S. Bolintineanu, W.M. Brown, P.S. Crozier, P.J.i.t. Veld, A. Kohlmeyer, S.G. Moore, T.D. Nguyen, R. Shan, M.J. Stevens, J. Tranchida, C. Trott, S.J. Plimpton, LAMMPS-a flexible simulation tool for particle-based materials modeling at the atomic, meso, and continuum scales, Computer Physics Communications, 271 (2022) 108171.

[46] S. Sengupta, S. Karmakar, C. Dasgupta, S. Sastry, Breakdown of the Stokes-Einstein relation in two, three, and four dimensions, J Chem Phys, 138 (2013) 12A548.

[47] S. Pan, Z.W. Wu, W.H. Wang, M.Z. Li, L. Xu, Structural origin of fractional Stokes-Einstein

relation in glass-forming liquids, Sci Rep, 7 (2017) 39938.

[48] G. Tarjus, D. Kivelson, Breakdown of the Stokes–Einstein relation in supercooled liquids, The Journal of Chemical Physics, 103 (1995) 3071-3073.

[49] S. Pan, Z. Wu, W. Wang, M. Li, L. Xu, Structural origin of fractional Stokes-Einstein relation in glass-forming liquids, Scientific reports, 7 (2017) 39938.

[50] S.R. Becker, P.H. Poole, F.W. Starr, Fractional Stokes-Einstein and Debye-Stokes-Einstein relations in a network-forming liquid, Physical review letters, 97 (2006) 055901.

[51] S. Corezzi, M. Beiner, H. Huth, K. Schröter, S. Capaccioli, R. Casalini, D. Fioretto, E. Donth, Two crossover regions in the dynamics of glass forming epoxy resins, J Chem Phys, 117 (2002) 2435-2448.

[52] W. Goetze, R. Schilling, Glass transitions and scaling laws within an alternative mode-coupling theory, Physical Review E, 91 (2015) 042117.

[53] A. Drozd-Rzoska, Universal behavior of the apparent fragility in ultraslow glass forming systems, Scientific Reports, 9 (2019) 6816.

[54] A. Drozd-Rzoska, S.J. Rzoska, S. Starzonek, New scaling paradigm for dynamics in glass-forming systems, Progress in Materials Science, 134 (2023).

[55] B. Wang, L. Luo, E. Guo, Y. Su, M. Wang, R.O. Ritchie, F. Dong, L. Wang, J. Guo, H. Fu, Nanometer-scale gradient atomic packing structure surrounding soft spots in metallic glasses, npj Computational Materials, 4 (2018) 41.

[56] M.L. Manning, A.J. Liu, Vibrational modes identify soft spots in a sheared disordered packing, Physical Review Letters, 107 (2011) 108302.

[57] Z. Zhang, J. Ding, E. Ma, Shear transformations in metallic glasses without excessive and predefinable defects, Proc Natl Acad Sci U S A, 119 (2022) e2213941119.

[58] Z.-Y. Yang, D. Wei, A. Zaccone, Y.-J. Wang, Machine-learning integrated glassy defect from an intricate configurational-thermodynamic-dynamic space, Physical Review B, 104 (2021) 064108.

[59] Z. Zhou, H. Wang, M. Li, Theoretical strength and prediction of structural defects in metallic glasses, Physical Review B, 100 (2019) 024109.

[60] C. Scalliet, L. Berthier, F. Zamponi, Nature of excitations and defects in structural glasses, Nature communications, 10 (2019) 5102.

[61] C. Chang, H. Zhang, R. Zhao, F. Li, P. Luo, M. Li, H. Bai, Liquid-like atoms in dense-packed solid glasses, Nature Materials, 21 (2022) 1240-1245.